\documentclass[preprint,nocopyrightspace]{sigplanconf}

\usepackage{epsfig}
\usepackage{endnotes}
\usepackage[utf8]{inputenc}
\usepackage[english]{babel}
\usepackage{authblk}
\usepackage{amsthm}
\usepackage{todonotes}
\usepackage{amsmath}
\usepackage{amssymb}
\usepackage{url}
\usepackage{listings}
\usepackage{balance}
\usepackage{epigraph}
\usepackage{fixltx2e}
\usepackage{graphicx}
\usepackage{algorithm}
\usepackage{algpseudocode}
\usepackage{caption}
\usepackage{color}
\usepackage{breakurl}
\usepackage{microtype}

\renewcommand\floatpagefraction{.9}
\renewcommand\topfraction{.9}

\usepackage[labelformat=simple]{subcaption}

\makeatletter
\let\OldStatex\Statex
\renewcommand{\Statex}[1][3]{%
  \setlength\@tempdima{\algorithmicindent}%
  \OldStatex\hskip\dimexpr#1\@tempdima\relax}
\makeatother

\theoremstyle{definition}

\theoremstyle{definition}

\theoremstyle{remark}

\usepackage[inline]{enumitem}

\begin{document}

\title{Dynamic Path Contraction for Distributed, Dynamic Dataflow Languages}

\authorinfo{Borja Arnau de Régil Basáñez\thanks{Work performed as a member of the BEAM Community project in the 2016 Google Summer of Code.}}
           {Complutense University of Madrid\\	
           Madrid, Spain}
           {bregil@ucm.es}
\authorinfo{Christopher S. Meiklejohn}
           {Université catholique de Louvain\\
           Louvain-la-Neuve, Belgium}
           {christopher.meiklejohn@uclouvain.be}

\maketitle

\begin{abstract}
We present a work in progress report on applying deforestation to distributed, dynamic dataflow programming models.  We propose a novel algorithm, dynamic path contraction, that applies and reverses optimizations to a distributed dataflow application as the program executes.  With this algorithm, data and control flow is tracked by the runtime system used to identify potential optimizations as the system is running.  We demonstrate and present preliminary results regarding this technique on an actor-based distributed programming model, Lasp, implemented on the Erlang virtual machine.
\end{abstract}



\section{Introduction}
In functional programming, the idea of intermediate values, or intermediate trees, refers to a value in a program whose only purpose is to serve as an intermediate step in a larger, and more
complex, computation. The process by which these values are removed is referred to as deforestation,
and is an important optimization process in the Glasgow Haskell Compiler~\cite{wadler1990deforestation,
coutts2007stream}.

This process, however, is static and is traditionally implemented as rewrite rules during the compilation phase. This is
problematic in distributed, dynamic dataflow languages, where available optimization opportunities may
present, or remove themselves at runtime. In addition, we want the system to adapt dynamically and
transparently during these situations: optimizations must be transparent to the user, and therefore
able to be reversed as the execution evolves.

The concept of dataflow analysis in compilers \cite{Rosen:1977:HDF:359842.359849} is well understood and
involves computing dependence relations in a program, and has multiple applications such as dead-code
elimination, liveness analysis, etc. Most optimization algorithms represent these relationships as a control
flow graph that can be analyzed and used as identifying potential targets in the application for optimization.

We present a work in progress report on a novel algorithm that combines these two ideas, called dynamic path
contraction, and apply it to a distributed, dynamic dataflow language. Dynamic path contraction composes two
main techniques: edge contraction and vertex cleaving. Edge contraction is used to remove intermediary results
that are only used as part of a larger computation. Vertex cleaving is used to reverse the effect of an
optimization: for instance, if a user decides to read an intermediary value that had been removed for optimization
purposes. We present an implementation of this algorithm and a preliminary evaluation performed using a
distributed, dynamic dataflow programming environment, called Lasp.

\begin{figure}[h]
  \begin{center}
    \noindent
    \includegraphics[scale=0.39]{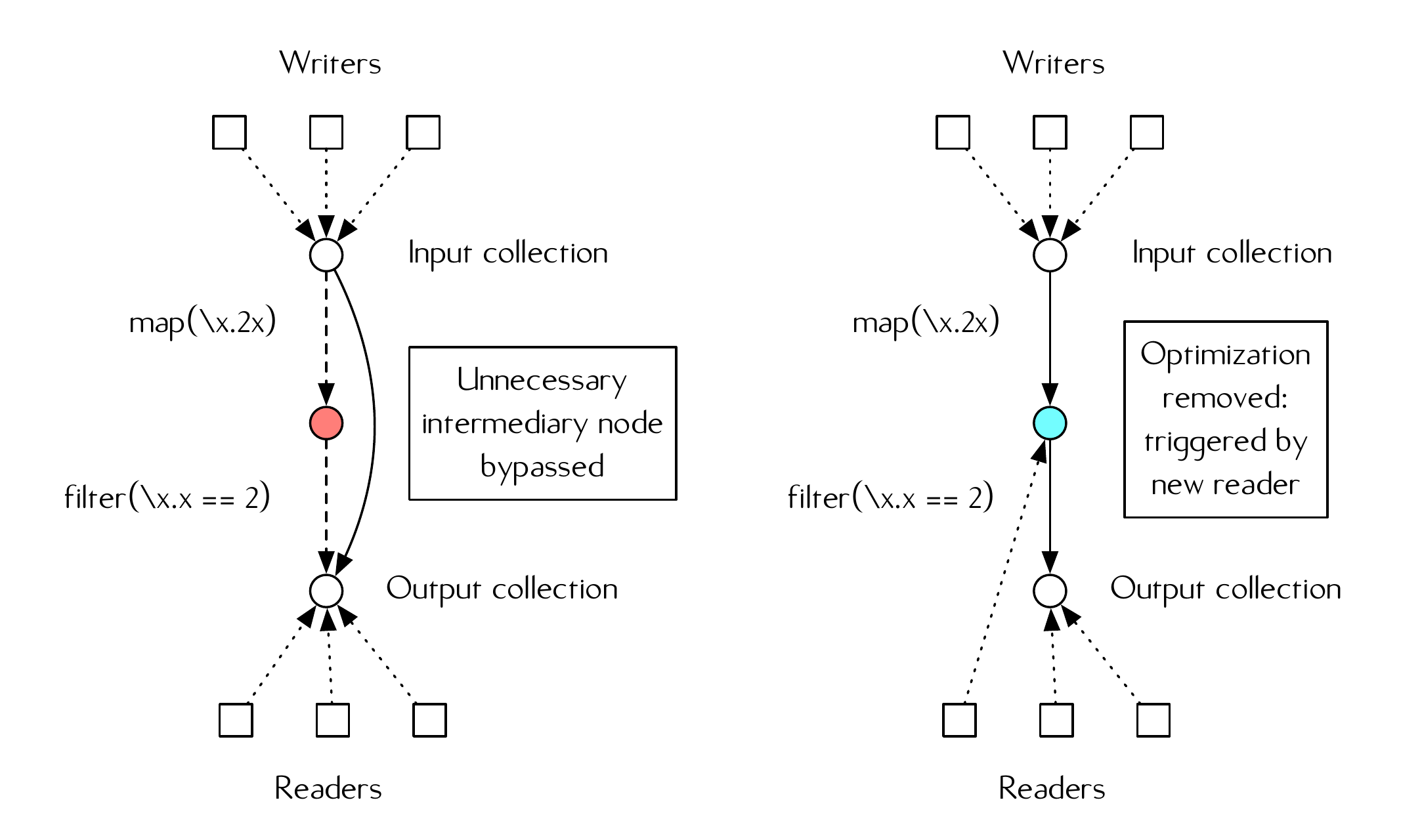}
  \end{center}
  \caption{Example application that contains two distributed input and output collections: the output is the result of applying a map and filter operation to the input.  When the intermediate value is not being actively read, the system can apply an optimization to remove the intermediate value.}
  \label{fig:example-application}
\end{figure}

\section{Background}
Lasp~\cite{meiklejohn2015lasp} is a distributed, fault-tolerant, dataflow programming model with a
prototype implementation in the actor-based language, Erlang.  It provides the user with high-level
functional programming capabilities for correctly building distributed systems; these capabilities
are implemented by a dataflow execution engine~\cite{meiklejohn2015implementation} modeled as a series
of cooperating actors that pass messages as data moves through the system. In Lasp, programs can be
modified as the system is executing either through the creation of new nodes in the dataflow topology or
through the introduction new clients to the system that will perform read or write operations on replicated collections.

Figure~\ref{fig:example-application} shows the topology of an example application written in Lasp.
Through this paper, we will only reference ``simple''\footnote{In this context, ``simple'' refers to acyclic
Lasp programs, as defined in \cite{meiklejohn2015lasp}.} Lasp applications, which present acyclic
topologies. In this example application, a distributed input collection is written to by multiple writers.
This collection is then used as the source of sequential map and filter operations that store the
result in the output collection. With this example, an intermediary node is used to store the output
of the map operation and provide the input to the filter operation. However, this is inefficient
because the intermediary results are not used directly, other than as a temporary storage
location.

Having intermediate values in a Lasp application reduces application performance for two reasons: 
\begin{enumerate*}
  \item the overhead of copying messages between actors in a shared-nothing actor model;
  \item and replication of values between instances of the runtime system that are not used other than as part of a larger computation.
\end{enumerate*}
Therefore, by applying dataflow analysis and dynamic path contraction, we get faster execution, lower distribution
latency, and potential bandwidth savings when a dataflow graph spans multiple nodes of a cluster.

Finally, as Lasp applications are dynamic and can be modified in real-time, any changes made as a result of our
optimizations have to be reversible, as they should be invisible to the user.

\section{Dynamic Path Contraction}
The main technique behind our optimization algorithm is called path contraction, a specific case
of edge contraction. In graph theory, edge contraction is an operation where an edge is removed
from a graph, while merging the two vertices that it previously joined. A path contraction operates
on the set of edges in a path, contracting them and leaving a single edge connecting the two endpoints
of the path. If there are any other edges connected to the intermediate vertices, they can be either
eliminated or arbitrarily connected to the endpoints.

To reverse any contractions introduced in a graph, we use what is called vertex splitting or cleaving,
which results in one vertex being split in two, and connecting them with a new edge.

In our algorithm, we represent an edge contraction, and therefore the merging of vertices, as the
``functional composition'' of the edges connecting them. To contract a path in our graph, we systematically compose
the intermediate edges until we arrive at a single edge connecting the endpoints in the path: our variation of vertex cleaving only operates on vertices that were previously merged by a contraction. The process
by which two edges are composed is explained in Section~\ref{edge_composition}. 

Our contribution, dynamic path contraction, focuses on the challenge of maintaining optimal performance
while the topology of the application changes. Our algorithm has two main components: a technique for
performing path contraction through edge composition, which removes unnecessary, intermediate values in
the program; and vertex cleaving, for reversing optimizations while the system is running, when an
intermediate value becomes necessary.

\subsection{Applications as Directed Acyclic Graphs}
Lasp applications are represented by a series of dataflow operations that connect collections with built-in functions that are both functional
(\texttt{map}, \texttt{filter}, \texttt{fold}) and set-theoretic (\texttt{union}, \texttt{product}, \texttt{intersection}). In our implementation, each function runs in their own separate process, which is identified by a unique identifier and implemented as an actor in Erlang, called a Lasp process.

We can generalize computations in the system in terms of three primitive operations: \textit{read}, \textit{transform}, and \textit{write}.
Any given function must \textit{read} from at least one input collection, apply a \textit{transform}-ation, and \textit{write} to an output collection.  For \texttt{map}, the process will \textit{read} from one input collection and apply a sequential map operation via \textit{transform} and \textit{write} the result into an output collection; for \texttt{union}, two input collections are used and one output collection produced.  We can use the \textit{read} and \textit{write} operations, in addition to variable declarations for input and output collections, to construct an acyclic dependency graph that represents all computation that is occurring in the system.

\begin{figure}[h]
  \begin{center}
    \noindent
    \includegraphics[scale=0.39]{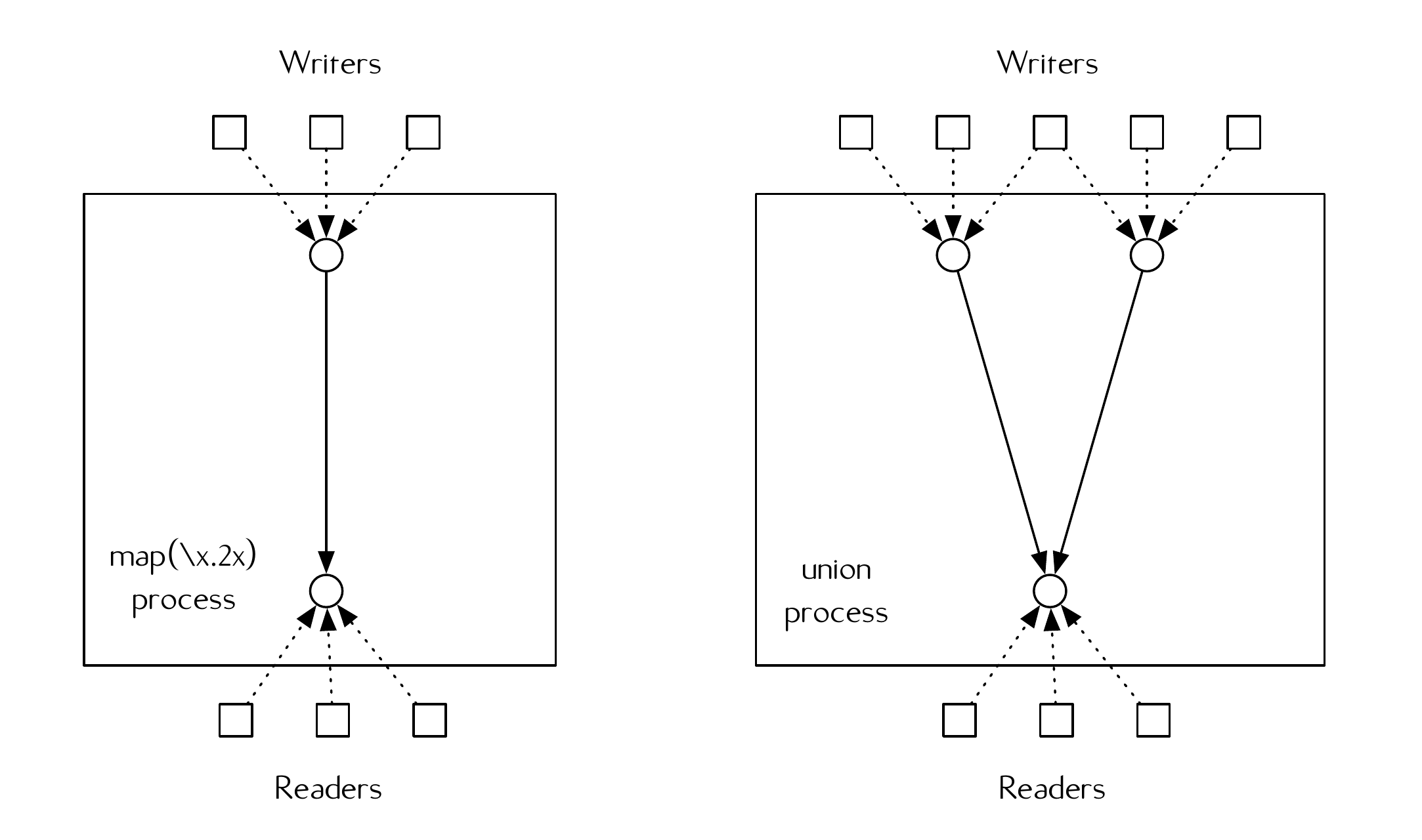}
  \end{center}
  \caption{Two examples of functional operations in Lasp: one is computing the map over a single collection; whereas, the other is computing the union of two collections.}
  \label{fig:flow-examples}
\end{figure}

Figure~\ref{fig:flow-examples} provides a visual depiction of both single arity functional operations in Lasp and two-arity set-theoretic operations.  In the first case, two vertices would exist: one for the input and one for the output collection.  There would be a single edge between these two vertices in a single process.  In the second case, three vertices would exit: one for each input and one for the output.  There would be two edges, each connecting one of the input collections to the output collection in a single process.

\subsection{Graph Construction}
\label{graph_construction}
We can construct a directed acyclic graph $G$ from a Lasp program as follows.  Our graph takes the form:
\begin{equation}
G = (V, E)	
\end{equation}

For each Lasp variable, we create a vertex $v$ for the variable.  We call this set of vertices $V$.
\begin{equation}
V = \{v_1, v_2, \ldots, v_n\}	
\end{equation}

For each Lasp process $p$, that connects one or more input vertices to an output vertex using read and write functions, we create an edge $e$, labeled by the process identifier $p$\footnote{We extend the normal ordered pair edge representation to support tracking a label.}.
\begin{equation}
E = \{(v_i, v_j, p_k) = e_1, \ldots, e_n\}	
\end{equation}

Whenever a user initiates a read or write operation on a variable from a user process, we generate a unique identifier to represent the user's process as a vertex and insert a new edge into the graph from the users process to the variable being read to or written from.

An edge from the unique identifier to a variable represents a process writing to a variable, whereas an edge from the variable to a unique identifier represents a process reading from a variable.
\begin{equation}
\label{eq:4}
\begin{split}
\mathrm{op}(\mathit{write}, p, v_i, (V, E)) : & \ v = \mathit{unique()} \\
								 	  & \ V' = V \cup \{ v \} \\
									  & \ E' = E \cup \{ (v, v_i, p) \} \\
									  & \ (V', E') \\							  
\mathrm{op}(\mathit{read}, p, v_i, (V, E)) : & \ v = \mathit{unique()} \\
								 	 & \ V' = V \cup \{ v \} \\
									 & \ E' = E \cup \{ (v_i, v, p) \}	\\
									 & \ (V', E')								  
\end{split}
\end{equation}

When processes terminate, their edges are removed from graph.

\subsection{Types of Vertices}
Among vertices that represent values, we distinguish two different types: necessary and unnecessary.  We say that a vertex is unnecessary if and only if it has an out-degree and in-degree of 1, and otherwise necessary.  Unnecessary vertices represent the intermediate values in our program.  We say that a given path is a \textbf{possible contraction path} if it connects two necessary vertices where all of the intermediate vertices are unnecessary.

\subsection{Edge Composition}
\label{edge_composition}
In Lasp, a function between two variables $v_i$ and $v_j$ can be defined as a triple,
\begin{equation}
f = \langle r_{v_i}, t_f, w_{v_j} \rangle
\end{equation}
where $r_{v_i}$ and $w_{v_j}$ are read and write functions, partially applied to $v_i$ and $v_j$, respectively.  The transform function\footnote{For functions that do not modify the input of the read function, the mathematical identity function is used as the transform function.}, $t_f$, modifies the return value of $r_{v_i}$ to produce the value that will be written by $w_{v_j}$.

Edge composition only operates on the sequence of edges that form a possible contraction path.
Given two edges, $(v_1, v_2)$ and $(v_2, v_3)$, representing two functions, $f$ and $g$,
\begin{equation}
\begin{split}
f = & \ \langle r_{v_1}, t_f, w_{v_2} \rangle \\
g = & \ \langle r_{v_2}, t_g, w_{v_3} \rangle
\end{split}
\end{equation}
their composition is defined as the composition of the functions they represent, resulting in a new
function $h$.
\begin{equation}
\begin{split}
h = g \circ f = \langle r_{v_1}, (t_g \circ t_f), w_{v_3} \rangle
\end{split}
\end{equation}

When created, this function will add a new edge $(v_1, v_3)$ to the graph, as described in equation \ref{eq:4}
in section \ref{graph_construction}. We call this edge the contraction edge. In addition, this procedure will
also terminate the processes containing the original functions $f$ and $g$, therefore removing the edges they
represent from the graph.

In the resulting graph representation, some of the vertices might be disconnected. These vertices represent
variables that are not used by the program, and are called contracted vertices.

\subsection{Vertex Cleaving}
\label{cleavingtheory}
Given the dynamic and distributed nature of the system, values that were previously safe to remove with
a path contraction may become necessary at some later point in the execution. This may happen if the user reads or writes to an unnecessary
vertex, or if a node in the topology got partitioned before a contraction and rejoined the cluster afterwards.
Therefore, any contractions we perform must be reversible. We call this process vertex cleaving\footnote{Specifically, vertex cleaving is the reverse operation of vertex contraction, the more general
version of edge contraction.}.

As described in the previous section, when two edges are composed together, we generate a new edge and
associated Lasp process, and remove the original edges from the graph. To be able to restore the graph to
its initial state, before a composition, we perform a ``soft-deletion'' of the old edges, by storing the
triples from the functions that represent those edges. In addition, vertices that get disconnected are
tagged with the unique identifier of the contraction edge.

When a contracted vertex becomes necessary, we terminate the process identified by its tag, thus removing
the associated edge from the graph. After that, we recreate the original functions, and edges, by retrieving
the stored triples. After a vertex cleaving is completed, we achieve a graph representing an identical topology
to the one present before applying path contraction on the affected vertex.

\section{Implementation}
We have implemented a version of dynamic path contraction in Lasp, a dynamic, distributed dataflow language written for the Erlang virtual machine.
We begin with an overview of how we construct our graph representation, followed by an exposition of the details of both path contraction and vertex cleaving.

\subsection{Lasp Processes and Process Supervision}

In Erlang, processes can monitor and link themselves to other processes. Both operations offer a degree of fault tolerance by letting us specify how the program should react when a process dies or is killed.  Supervision trees, or supervisors, allow us to specify policies that are used to control process restarts and are built on top of primitives like monitors, that allow us to identify when actors fail.

When a Lasp program starts, we create a graph actor, that will maintain the graph representation of the system,
and perform path contractions and vertex cleavings as the program executes. To build our dependency graph, we track all Lasp processes and its dependencies at run-time. We made
all processes register themselves, by calling the graph actor, as part of their creation step. 

When a Lasp process is registered, the system stores a cache of each input and output value, as well as their read, transform
and write functions. The graph actor will, in turn, monitor the execution of these processes. If one dies or
gets restarted by a supervisor, the graph actor will be notified. When this happens, it will update the graph
by removing the appropriate edges.

\subsection{Dynamic Path Contraction in Lasp}
\subsubsection*{Path Contraction}
The first step to perform a path contraction is to identify the possible contraction paths. We call
an optimization pass to the process of searching the graph for these paths. In our current implementation,
optimization passes are scheduled to happen at regular intervals.

To find possible contraction paths, we traverse the graph in topological order, keeping track of all
the visited vertices. When we find an unnecessary vertex, we extend the search both ``upwards'' and ``downwards'',
collecting all the vertices, until we find necessary vertices, denoting the endpoints of a possible
contraction path: this process continues until we completely traverse the graph.

Once completed, we contract all possible contraction paths that were collected. To create a new Lasp
process that connects the endpoints of the path, we must generate new read, transform and write functions.
To define the new read, write and transform functions, we select the read function defined at the first
edge in the path, the write function defined at the last edge in the path, and perform function
composition of all intermediate transform functions along the path. This three-tuple defines a new
Lasp process that is atomically started by the system while the intermediary edges are removed.

\subsubsection*{Vertex Cleaving}
\label{cleavingimplementation}
Our current implementation of vertex cleaving mostly follows what was already outlined in section
\ref{cleavingtheory}. One thing to note is that vertex cleavings will remove a contraction
optimization for an entire path even if it was only necessary for a single vertex. This makes
it less efficient than it could be. In section \ref{conclusions} we elaborate on how to improve this.




\section{Evaluation}
We now analyze the current implementation of dynamic path contraction by looking at the
performance of different Lasp applications. We have implemented a compiler for a limited
subset of SQL that transforms queries into Lasp applications. To gauge the effectiveness
of our algorithm, we look at several examples and measure the end-to-end distribution latency
of values as updates propagate.

For each test, we first build an application topology. We then identify the longest path,
and measure the time it takes for an update on one endpoint to reach the other.
We also analyze two variants: an optimized version (\textit{contraction}), where all the possible
contraction paths have been optimized and an unoptimized version (\textit{no contraction}).

\subsection{Contraction Opportunities}
Given the nature of our optimization, its impact on performance depends on two program
characteristics.

\begin{itemize}

  \item{\textbf{The number of intermediate values in the program.}} This is the main and most important
        factor that influences the effectiveness of the contractions. Programs that have
        a great number of intermediate values present more contraction opportunities, and
        therefore are more affected by the optimization.

  \item{\textbf{Dynamism of the system.}} A program that presents a very dynamic graph representation
        is less amenable to optimizations. As values selected as unnecessary become necessary
        (and vice versa), our algorithm has to contract and cleave new sections of the graph.

\end{itemize}

\subsection{Simple Contractions}
We first apply our implementation to a simple application, consisting of 5 vertices along
a single path, creating three intermediate vertices. In this simple program no mutation is
performed along the path, measuring only value distribution latency. Intermediate values
were read every 10 iterations. We update the value of the first vertex in the path, and
measure the end-to-end latency of the result to appear in the last vertex.

\begin{figure}[h]
  \begin{center}
    \noindent
    \includegraphics[scale=0.6]{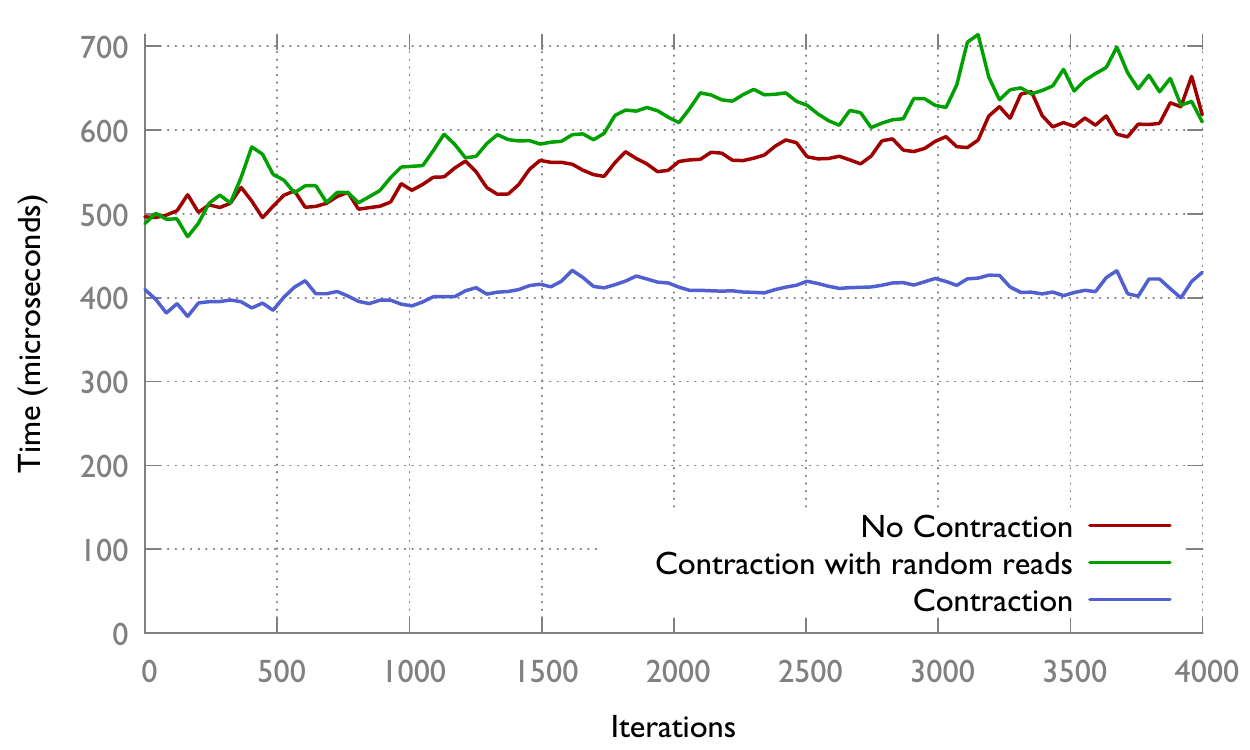}
  \end{center}
  \caption{Latency along a single contraction path where three vertices were removed.}
  \label{fig:simple-contraction}
\end{figure}

For this example, we also show an optimized version where we perform a read from a contracted
vertex at random\footnote{Following a uniform distribution.} (\textit{contraction with random reads}),
in order to force the algorithm to perform a vertex cleaving. Figure~\ref{fig:simple-contraction} shows
the difference in latency for the three variants; the x-axis shows the number of updates, while the y-axis
shows latency (in microseconds, lower is better).

With a stable graph representation, path contraction results in a 25\% decrease in median
distribution latency. With a dynamic graph representation, we see a 5\% increase in
distribution latency. In our current implementation, we pause the distribution of values while
performing vertex cleaving, which may explain this increase in latency. In Section~\ref{conclusions},	
we provide indications on how to further improve these results.

\subsection{Dynamic Path Contraction for SQL}
\begin{figure}[h]
  \begin{center}
    \noindent
    \includegraphics[scale=0.38]{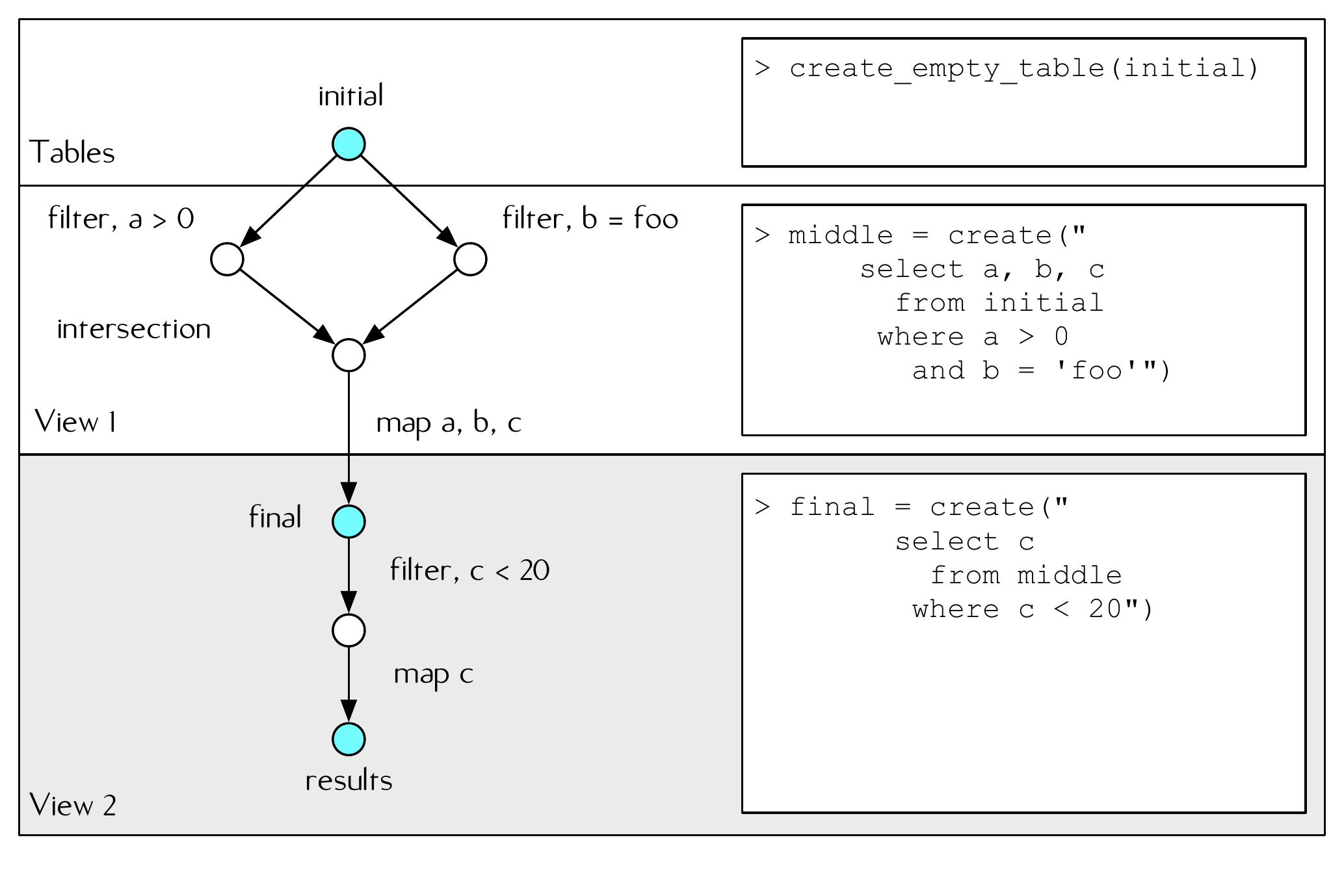}
  \end{center}
  \caption{Lasp can support a minimal subset of SQL.}
  \label{fig:sql-join-example}
\end{figure}

We now show how our algorithm can improve distribution latency in Lasp applications generated from SQL queries. In this example we perform two queries with two composed views. Figure~\ref{fig:sql-join-example} shows the Lasp program and the resulting dataflow graph.

\begin{figure}[h]
  \centering
  \noindent
    \includegraphics[scale=0.6]{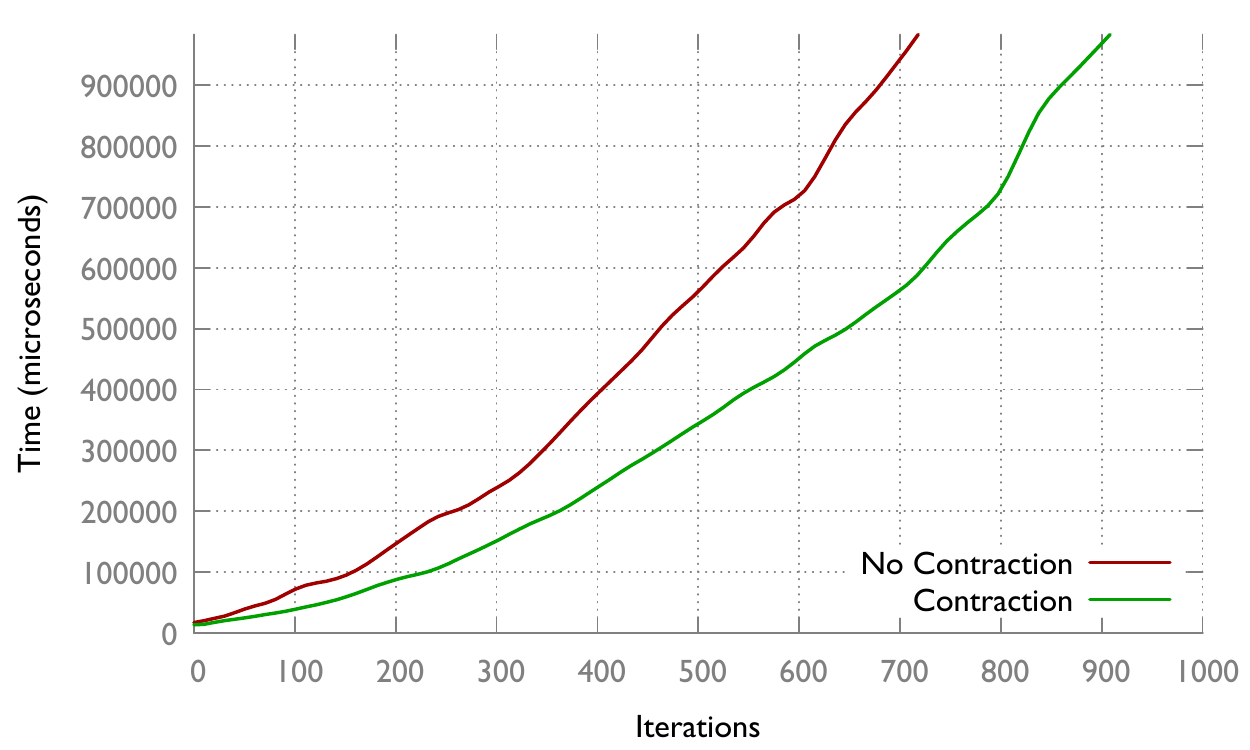}
  \caption{Distribution latency for Figure~\ref{fig:sql-join-example}.}
  \label{fig:sql-join-graph}
\end{figure}

We populate the first collection with a number of rows, modeled as individual dictionaries each with a unique key inside of a collection and measured the end-to-end latency of the result
to appear in the final collection. The results are shown in Figure~\ref{fig:sql-join-graph}. By
applying dynamic path contraction, we get a 38\% decrease in median distribution latency\footnote{It is important to note that Lasp does not have a mechanism for incremental view maintenance, so, as each insert operation is performed the full state is copied along the entire path.}.

\section{Conclusions and Future Work}
\label{conclusions}
We have presented dynamic path contraction, a deforestation technique for
distributed dataflow programming models. Dynamic path contraction combines
different approaches from deforestation and control flow analysis in order
to remove intermediate values at runtime, achieving reduced distribution
latency while adapting itself to changes in the application topology.

We have identified several points where we could improve our current implementation.

Contractions only remove intermediate values between unary functions, reducing the
number of optimization opportunities we can detect. We think that extending the
current implementation to support functions of arbitrary arity is possible,
and should result in greater efficiency.

Our current cleaving strategy involves removing the contraction optimization
for an entire path even if only one vertex becomes necessary. We think its
possible to selectively cleave parts of contracted path, in order to make
the cleaving process less wasteful, thus removing the current latency overhead
for programs with a high level of dynamism.

\balance

\bibliographystyle{abbrvnat}
\bibliography{dynamic-path-contraction}

\end{document}